\documentclass{lmcs} 
\pdfoutput=1

\usepackage{lastpage}
\lmcsdoi{18}{1}{20}
\lmcsheading{}{\pageref{LastPage}}{}{}%
{Dec.~15,~2017}{Jan.~21,~2022}{}

\usepackage[utf8]{inputenc}
\usepackage{graphics,graphicx,epsfig} 
\usepackage{tikz}
\usetikzlibrary{arrows}
\usepackage{url}
\usepackage{latexsym}
\usepackage{amssymb,amsfonts,amsmath}

\newcommand{\eq}{\leftrightarrow}
\newcommand{\Eq}{\Leftrightarrow}
\newcommand{\imp}{\rightarrow}

\newcommand{\et}{\wedge}

\newcommand{\proves}{\vdash}

\newcommand{\Dia}{\Diamond}
\newcommand{\dia}[1]{\langle #1 \rangle}

\newcommand{\M}{\hat{K}}

\renewcommand{\phi}{\varphi}
\newcommand{\union}{\cup}
\newcommand{\Union}{\bigcup}
\newcommand{\inter}{\cap}

\newcommand{\bisim}{{\raisebox{.3ex}[0mm][0mm]{\ensuremath{\medspace \underline{\! \leftrightarrow\!}\medspace}}}}
\newcommand{\bisrel}{\ensuremath{\mathfrak{R}}}

\newcommand{\weg}[1]{}

\newcommand{\lbr}{[\![}
\newcommand{\rbr}{]\!]}
\newcommand{\I}[1]{\lbr #1 \rbr} 
\newcommand{\II}[1]{\lbr #1 \rbr} 

\newcommand{\ov}{\overline}

\newcommand{\Formulas}{{\mathcal L}}

\newcommand{\langu}{\Formulas}

\newcommand{\state}{s}

\newcommand{\stateb}{t}

\newcommand{\Atoms}{P}

\newcommand{\atom}{p}

\newcommand{\Agents}{A}

\newcommand{\agent}{a}

\newcommand{\Domain}{{\mathcal D}}
\newcommand{\domain}{\Domain}

\newcommand{\Nat}{\mathbb N}
\newcommand{\Naturals}{\Nat}

\newcommand{\lang}{\langu}

\newcommand{\langpl}{\ensuremath{\lang_{\mathit{pl}}}}
\newcommand{\langel}{\ensuremath{\lang_{\mathit{el}}}}

\newcommand{\langpal}{\ensuremath{\lang_{\mathit{pal}}}}
\newcommand{\langapal}{\ensuremath{\lang_{\mathit{apal}}}}
\newcommand{\langbapal}{\ensuremath{\lang_{\mathit{bapal}}}}
\newcommand{\langaanf}{\ensuremath{\lang_{\mathit{aanf}}}}

\newcommand{\logicel}{\ensuremath{\mathit{EL}}}
\newcommand{\logicpal}{\ensuremath{\mathit{PAL}}}
\newcommand{\logicapal}{\ensuremath{\mathit{APAL}}}
\newcommand{\logicbapal}{\ensuremath{\mathit{BAPAL}}}
\newcommand{\logicpapal}{\ensuremath{\mathit{APAL+}}}
\newcommand{\logicgal}{\ensuremath{\mathit{GAL}}}

\newcommand{\knows}{K}

\newcommand{\var}{\mathit{var}}

\newcommand{\axiombapal}{\ensuremath{\text{\pmb{\sc bapal}}}}

\usepackage{rotating}

\begin{document}

\title{Quantifying over Boolean announcements}


\author[H.~van Ditmarsch]{Hans van Ditmarsch\rsuper{a}}	
\address{Open University, Heerlen, the Netherlands}	
\email{hans.vanditmarsch@ou.nl}  

\author[T.~French]{Tim French\rsuper{b}}	
\address{University of Western Australia, Perth, Australia}	
\email{tim.french@uwa.edu.au}  


\begin{abstract}
Various extensions of public announcement logic have been proposed with quantification over announcements. The best-known extension is called arbitrary public announcement logic, \logicapal. It contains a primitive language construct $\Box\phi$ intuitively expressing that ``after every public announcement of a formula, formula $\phi$ is true''. The logic \logicapal{} is undecidable and it has an infinitary axiomatization. Now consider restricting the \logicapal{} quantification to public announcements of Boolean formulas only, such that $\Box\phi$ intuitively expresses that ``after every public announcement of a Boolean formula, formula $\phi$ is true''. This logic can therefore be called {\em Boolean arbitrary public announcement logic}, \logicbapal. The logic \logicbapal{} is the subject of this work. Unlike \logicapal{} it has a finitary axiomatization. Also, \logicbapal{} is not at least as expressive as \logicapal. A further claim that \logicbapal{} is decidable is deferred to a companion paper.
\end{abstract}

\maketitle

\section{Introduction} \label{sec.one}

Public announcement logic (\logicpal{})~\cite{gerbrandyetal:1997,plaza:1989} extends
epistemic logic with operators for reasoning about the effects of specific
public announcements. The formula $[\psi] \phi$ means that ``$\phi$
is true after the truthful announcement of $\psi$''. This means that, when interpreted in a Kripke model with designated state, after submodel restriction to the states where $\phi$ is true (this includes the designated state, `truthful' here means true), $\psi$ is true in that restriction. Arbitrary public
announcement logic (\logicapal{})~\cite{balbianietal:2008} augments this with
operators for quantifying over public announcements. The formula
$\Box \phi$ means that ``$\phi$ is true after the truthful announcement of
any formula that does not contain $\Box$''. 

Quantifying over the communication of information as in
\logicapal{} has applications to epistemic protocol synthesis, where
we wish to achieve epistemic goals by communicating information to agents, but
where we do not know of a specific protocol that will achieve the goal, and where we may not
even know if such a protocol exists. In principle, synthesis problems can be solved
by specifying them as formulas in the logic, and applying model-checking or
satisfiability procedures. However in the case of \logicapal{}, while there is
a \textsc{PSpace}-complete model-checking procedure~\cite{agotnesetal.jal:2010}, the 
satisfiability problem is undecidable in the presence of multiple agents~\cite{frenchetal:2008}. 

The quest for a decidable version of public announcement logic with quantification has been going on for a while. Group announcement logic and coalition announcement logic, that are logics with quantifiers over public information change that are similar to \logicapal, are also undecidable~\cite{agotnesetal:2016}. Whereas the `mental model' arbitrary public announcement logic of~\cite{charrieretal:2015} is decidable. Yet other dynamic epistemic logics have more generalized quantification, namely over non-public information change. Of those, arbitrary arrow update logic~\cite{hvdetal.undecidable:2017} is undecidable, whereas refinement modal logic~\cite{bozzellietal.inf:2014} and arbitrary action model logic~\cite{hales2013arbitrary} are decidable. The above-mentioned~\cite{charrieretal:2015}, wherein a decidable arbitrary public announcement logic is presented, is an interesting case. Decidability is achieved by restricting epistemic modalities, while retaining arbitrary announcements (of formulas containing such modalities). These special modalities are not labelled with an (abstract set of) agents, but with programs using propositional variables. This severely constrains the relational (`Kripke') models (possibly) satisfying the formulas bound by the epistemic modalities, which is how decidability can then be obtained for this logic. Instead, in the logic that we will propose, we do not restrict the epistemic modalities, but restrict the quantification over the announcements, the dynamic modalities. 

We propose a multi-agent epistemic logic with public announcements and with quantification over public announcements of Boolean formulas (so-called Boolean announcements). We call this {\em Boolean arbitrary public announcement logic} (\logicbapal). It is therefore a version of \logicapal: as said, in \logicapal{} we allow quantification over any quantifier-free ($\Box$-free) formulas, including formulas with announcement modalities and knowledge modalities. For this logic we obtain the following results. 
%
\begin{quote} {\it Unlike \logicapal, \logicbapal{} has a (complete) {\bf finitary axiomatization}.} \end{quote} 
For \logicapal{} only an infinitary axiomatization is known~\cite{balbianietal:2015}, although it has not been proved that a finitary axiomatization cannot exist~\cite{Kuijer17}. A finitary axiomatization for a logic much like \logicapal, called `\logicapal{} with memory', was proposed in~\cite{BaltagOS18}. Its structures are topological and its semantics allow to refer to states prior to an announcement (it is history-based). As non-bisimilar states can become bisimilar after an announcement, this semantics makes it possible to continue to distinguish such states; which is the basis for the soundness of their complete axiomatization. Their $\Box$ quantifies over epistemic formulas, which in their setting is different from quantifying over formulas that may also contain announcements (the $\Box$-free formulas mentioned above). The semantics are therefore different from that of \logicapal{} and also different from those of \logicbapal, with quantification over Booleans. It is an open question if `\logicapal{} with memory' and \logicapal{} have the same validities.

\begin{quote} \logicbapal{} is {\bf not at least as expressive} as \logicapal. \end{quote}
Also, \logicbapal{} is more expressive than \logicpal, which can be shown just as for \logicapal. An open question remains whether \logicapal{} is not at least as expressive as \logicbapal. 

We further claim that:
\begin{quote} {\it Unlike \logicapal, \logicbapal{} is {\bf decidable}.} \end{quote} 
As most such logics with quantification over information change are undecidable, this seems remarkable. That result in not reported in this work but in a companion paper in progress.

There seems to be many applications, in particular in planning, wherein it makes sense only to consider quantifications over Booleans. 

Consider cards cryptography wherein two communicating agents (the principals) attempt to learn the card deal without other players (eavesdroppers) learning the card deal, or even something stronger, such as not learning the ownership of any single card other than their own~\cite{fischeretal:1996,hvd.studlog:2003,cordonetal.tcs:2013}. When modelling initial uncertainty about a stack of cards being dealt over a finite set of players, single states in such models can be uniquely identified with a card deal. Therefore, public announcements restricting such models correspond to Booleans. For example, let there be three cards $0,1,2$ and three players Alice, Bob, and Eve, and suppose that Alice announces (truthfully) that she holds card $0$. This corresponds to the public announcement of (some elementary Boolean representation of) two card deals, namely the one wherein, additionally, Bob holds 1 and Eve holds 2, and the one wherein Bob holds 2 and Eve holds 1. 

As another example, consider multi-agent planning for publicly observable sensing actions under uncertainty~\cite{Levesque96,hvdetal.sitcalc:2011,bolanderetal:2011,CharrierHLMS16}: given multiple agents that are uncertain about a number of system parameters (lights, switches, temperature settings) they may be informed, or they may be informing each other, about their observations of the state of the light. Or they may be planning to make such observations, and contingent on the  outcome of such observations take further action.

\medskip

We close the introduction with an outline of the content of this work. In Section~\ref{sec.two} we define the logical language and semantics of \logicbapal{} and in the subsequent Section~\ref{sec.twothree} we give various results for this semantics that will be used in later sections. Section~\ref{sec.three} is on the expressivity of \logicbapal. Section~\ref{sec.four} presents the complete axiomatization. 

\section{Boolean arbitrary public announcement logic} \label{sec.two}

Given are a countable (finite or countably infinite) set of {\em agents} $\Agents$ and a countably infinite set of {\em propositional variables} $\Atoms$ (a.k.a.\ {\em atoms}, or {\em variables}).

\subsection{Syntax} We start with defining the logical language and some crucial syntactic notions. 
\begin{defi}[Language] 
The language of Boolean arbitrary public announcement logic is defined as follows,  where $\agent\in\Agents$ and $\atom\in\Atoms$. 
\[ \langbapal(\Agents,\Atoms) \ \ni \ \phi ::= \atom \ | \ \neg \phi \ | \ (\phi \et \phi) \ | \ K_a \phi \ | \ [\phi]\phi \ | \ \Box \phi \] 
\end{defi}
Other propositional connectives are defined by abbreviation. For $K_a \phi$ read `agent $a$ knows $\phi$'. For $\Box \phi$, read `after any Boolean announcement, $\phi$ (is true)'. For $[\phi]\psi$, read `after public announcement of $\phi$, $\psi$'. The dual modalities are defined by abbreviation: $\hat{K}_a\phi := \neg K_a \neg \phi$, $\dia{\phi}\psi := \neg [\phi] \neg \psi$, and $\Dia\phi := \neg \Box \neg \phi$. Unless ambiguity results we often omit one or both of the parameters $\Agents$ and $\Atoms$ in $\langbapal(\Agents,\Atoms)$, and write $\langbapal(\Atoms)$ or $\langbapal$. Unless ambiguity results we often omit parentheses occurring in formulas. Formulas are denoted $\phi,\psi$, possibly primed as in $\phi',\phi'',\dots, \psi',\dots$

We also distinguish the language $\langel$ of \emph{epistemic logic} (without the constructs $[\phi]\phi$ and $\Box\phi$) and the language $\langpl$ of \emph{propositional logic} (without additionally the construct $K_a \phi$), also known as the \emph{Booleans}. Booleans are denoted $\phi_0, \psi_0$, etc. 

The set of propositional variables that occur in a given formula $\phi$ is denoted $\var(\phi)$ (where one that does not occur in $\phi$ is called a {\em fresh variable}), its {\em modal depth} $d(\phi)$ is the maximum nesting of $K_a$ modalities, and its {\em quantifier depth} $D(\phi)$ is the maximum nesting of $\Box$ modalities. These notions are inductively defined as follows.
\begin{itemize}
\item $\var(p) = \{p\}$, $\var(\neg\phi) = \var(K_a \phi) = \var(\Box \phi) = \var(\phi)$, $\var(\phi\et\psi) = \var([\phi]\psi) = \var(\phi) \union \var(\psi)$; \item $D(p) = 0$, $D(\neg\phi) = D(K_a \phi) = D(\phi)$, $D(\phi\et\psi) = D([\phi]\psi) = \max \{D(\phi), D(\psi)\}$, $D(\Box \phi) = D(\phi)+1$; \item $d(p) = 0$, $d(\neg\phi) = d(\Box \phi) = d(\phi)$, $d(\phi\et\psi) = \max \{d(\phi), d(\psi)\}$, $d([\phi]\psi) = d(\phi)+d(\psi)$, $d(K_a \phi) = d(\phi)+1$.
\end{itemize}

{\em Arbitrary announcement normal form} is a syntactic restriction of $\langbapal$ that pairs all public announcements with arbitrary Boolean announcement operators. It plays a role in the decidability proof. We will show that any formula in \langbapal{} is equivalent to one in \langaanf.
\begin{defi}[Arbitrary announcement normal form] \label{def:aanf}
The language fragment $\langaanf$ is defined by the following syntax, where $\agent\in\Agents$ and $\atom\in\Atoms$.
$$\phi ::= \atom \ | \ \neg\phi\ |\ (\phi\et\phi) \ |\ \knows_\agent\phi\ |\ [\phi]\Box\phi$$
\end{defi}

We now define necessity forms and possibility forms. Necessity forms are used in derivation rules in the proof system.
\begin{defi}[Necessity form]
Consider a new symbol $\sharp$. The {\em necessity forms} are defined inductively as follows, where $\phi \in \langbapal$ and $a \in \Agents$.
    $$\psi(\sharp) ::= \sharp \mid (\phi \imp \psi(\sharp)) \mid K_a \psi(\sharp) \mid [\phi]\psi(\sharp)$$ \end{defi} By induction on the necessity form $\psi(\sharp)$ the reader may easily verify that each $\psi(\sharp)$ contains a unique occurrence of the symbol $\sharp$. If $\psi(\sharp)$ is a necessity form and $\phi \in \langbapal$, then $\psi(\phi)$ is $\psi(\sharp)[\phi/\sharp]$ (the substitution of $\sharp$ in $\psi(\sharp)$ by $\phi$), where we note that $\psi(\phi) \in \langbapal$. A {\em possibility form} is the dual of a necessity form. They are therefore defined as: \[ \psi\{\sharp\} ::= \sharp \mid (\phi \et \psi\{\sharp\}) \mid \M_a \psi\{\sharp\} \mid \dia{\phi}\psi\{\sharp\} \] 
Similarly to above, notation $\psi\{\phi\}$ means that $\sharp$ is substituted by $\phi$ in $\psi\{\sharp\}$. 

Given necessity form $\psi(\sharp)$, let $\overline{\psi}\{\sharp\} = t(\psi(\sharp))$ be obtained by defining inductively: $t(\sharp) = \sharp$, $t(\phi \imp \psi(\sharp)) = \phi \et t(\psi(\sharp))$, $t(K_a \psi(\sharp)) = \M_a t(\psi(\sharp))$ and $t([\phi]\psi(\sharp)) = \dia{\phi}t(\psi(\sharp))$. Note that $\overline{\psi}\{\sharp\}$ is indeed a possibility form. We will later show that $\neg \psi(\neg \phi)$ is equivalent to $\overline{\psi}\{\phi\}$.

\subsection{Structures} 
We consider the following structures and structural notions in this work.
\begin{defi}[Model]
An {\em (epistemic) model} $M = (S, \sim, V )$ consists of a non-empty {\em domain} $S$ (or $\domain(M)$) of {\em states} (or `worlds'), an {\em accessibility function} $\sim: \Agents \imp {\mathcal P}(S \times S)$, where each $\sim_\agent$ is an equivalence relation, and a {\em valuation} $V: \Atoms \imp {\mathcal P}(S)$, where each $V(\atom)$ represents the set of states where $\atom$ is true. For $\state \in S$, a pair $(M,s)$, for which we write $M_s$, is a {\em pointed (epistemic) model}. \end{defi} 
We will abuse the language and also call $M_s$ a model. We will occasionally use the following disambiguating notation: if $M$ is a model, $S^M$ is its domain, $\sim^M_a$ the accessibility relation for an agent $a$, and $V^M$ its valuation.
\begin{defi}[Bisimulation]\label{bisimulation}
    Let $M = (S,\sim,V)$ and $M' = (S',\sim',V')$
    be epistemic models. 
    A non-empty relation $\bisrel \subseteq S \times S'$
    is a {\em bisimulation} if for every 
    $(s, s') \in \bisrel$,
    $p \in P$, and
    $a \in A$ 
    the conditions {\bf atoms}, {\bf forth} and {\bf back} hold.
\begin{itemize}
\item    {\bf atoms}: 
    $s \in V(p)$ iff $s' \in V'(p)$.

\item    {\bf forth}: 
    for every $t \sim_{a} s$ 
    there exists $t' \sim'_{a} s'$
    such that $(t, t') \in \bisrel$.

\item    {\bf back}: 
    for every $t' \sim'_{a} s'$
    there exists $t \sim_{a} s$ 
    such that $(t, t') \in \bisrel$.
\end{itemize}
If there exists a bisimulation $\bisrel$ between $M$ and $M'$ such that $(s, s') \in \bisrel$, then 
    $M_{s}$ and $M'_{s'}$ are
    {\em bisimilar}, notation  
    $M_{s} \bisim M'_{s'}$ (or $\bisrel: M_{s} \bisim M'_{s'}$, to be explicit about the bisimulation).

Let $Q \subseteq P$. A relation $\bisrel$ between $M$ and $M'$ satisfying {\bf atoms} for all $p \in Q$, and {\bf forth} and {\bf back}, is a {\em $Q$-bisimulation} (a bisimulation {\em restricted} to $Q$). The notation for $Q$-restricted bisimilarity is $\bisim_Q$.
\end{defi}

\noindent The notion of $n$-bisimulation, for $n \in \Naturals$, is given by defining  relations $\bisrel^0 \supseteq \dots \supseteq \bisrel^n$.
\begin{defi}[$n$-Bisimulation] \label{n-bisimulation}
    Let $M = (S,\sim,V)$ and $M' = (S',\sim',V')$
    be epistemic models, and let $n \in \Naturals$. 
    A non-empty relation $\bisrel^0 \subseteq S \times S'$ 
    is a {\em $0$-bisimulation} if {\bf atoms} holds for pair $(s,s') \in\bisrel$. Then, a non-empty relation $\bisrel^{n+1} \subseteq S \times S'$ 
    is a {\em $(n+1)$-bisimulation} if for all $p \in P$ and $a \in A$:
\begin{itemize}
%
\item    {\bf $(n+1)$-forth}: 
    for every $t \sim_{a} s$ 
    there exists $t' \sim'_{a} s'$
    such that $(t,t') \in \bisrel^n$;
\item    {\bf $(n+1)$-back}: 
    for every $t' \sim'_a s'$
    there exists $t \sim_{a} s$ 
    such that $(t,t') \in \bisrel^n$.
\end{itemize}
Similarly to $Q$-bisimulations we define {\em $Q$-$n$-bisimulations}, wherein {\bf atoms} is only required for $p \in Q \subseteq P$; $n$-bisimilarity is denoted $M_s \bisim^n M'_{s'}$, and $Q$-$n$-bisimilarity is denoted $M_s \bisim^n_Q M'_{s'}$.
\end{defi}

\subsection{Semantics}
We continue with the semantics of our logic.
\begin{defi}[Semantics] \label{def.truthlyingpub}
The interpretation of formulas in $\langbapal$ on epistemic models is defined by induction on formulas.

Assume an epistemic model $M = (S, \sim, V )$, and $s \in S$.  
\[ \begin{array}{lcl}
M_s \models \atom &\mbox{iff} & \state \in V(p) \\ 
M_s \models \neg \phi &\mbox{iff} & M_s \not \models \phi \\ 
M_s \models \phi \et \psi &\mbox{iff} & M_s \models \phi  \text{ and } M_s \models \psi \\  
M_s \models K_\agent \phi &\mbox{iff} & \mbox{for all }  \stateb \in S: \state \sim_a \stateb \text{ implies } M_t  \models \phi \\  
M_s \models [\phi] \psi &\mbox{iff} & M_s \models \phi \text{ implies } M^\phi_s \models \psi \\
M_s \models \Box \psi & \mbox{iff} & \mbox{for all } \phi_0 \in \langpl : M_s \models [\phi_0] \psi
\end{array} \] 
where $\II{\phi}_M := \{ s\in S \mid M_s\models \phi\}$; and where epistemic model $M^\phi = (S', \sim', V')$ is such that: $S' = \II{\phi}_M$, ${\sim'_a} = {\sim_a} \inter (\II{\phi}_M \times \II{\phi}_M)$, and $V'(p) := V(p) \inter \II{\phi}_M$. For $(M^\phi)^\psi$ we may write $M^{\phi\psi}$. Formula $\phi$ is {\em valid on model $M$}, notation $M \models \phi$, if for all $s \in S$, $M_s \models \phi$. Formula $\phi$ is {\em valid}, notation $\models \phi$, if for all $M$, $M \models \phi$.
\end{defi} 
%
Given $M_s$ and $M'_{s'}$, if for all $\phi \in\langbapal$, $M_s \models \phi$ iff $M'_{s'} \models \phi$, we write $M_{s} \equiv M'_{s'}$. Similarly, if this holds for all $\phi$ with $d(\phi) \leq n$, we write $M_{s} \equiv^n M'_{s'}$, and if this holds for all $\phi$ with $\var(\phi) \in Q \subseteq P$, we write $M_s \equiv_Q M'_{s'}$.  


Note that the languages of $\logicapal$ and $\logicbapal$ are the same, but that their semantics are different. The only difference is the interpretation of $\Box \phi$: in $\logicapal$, this quantifies over the $\Box$-free fragment~\cite{balbianietal:2008}, so that, given the eliminability of public announcements from that fragment~\cite{plaza:1989}, this amounts to quantifying over formulas of epistemic logic:
\[ \begin{array}{lcl}
M_s \models \Box \psi & \mbox{iff} & \mbox{for all } \phi \in \langel : M_s \models [\phi] \psi \quad \quad \quad \quad \text{(\logicapal\ semantics of } \Box \phi)
\end{array} \] 
 
\section{Semantic results} \label{sec.twothree}
We continue with basic semantic results for the logic. They will be used in various of the later sections. Various well-known results for any dynamic epistemic logic with propositional quantification generalize straightforwardly to \logicbapal. 

\subsection{Bisimulation invariance}

We start with the bisimulation invariance of \logicbapal. This is shown as for \logicapal.

\begin{lem}\label{bisimulation-preserves}
    Let $M_{s}, N_{s'}$ be epistemic models. Then $M_{s} \bisim N_{s'}$ implies $M_{s} \equiv N_{s'}$.
\end{lem}
\begin{proof}
We prove that: for all $\phi\in\langbapal$, and for all $M_{s}, N_{s'}$: if $M_{s} \bisim N_{s'}$, then $M_s \models \phi$ iff $N_{s'} \models \phi$; from which the required follows by restricting the scope of $\phi$ to the consequent of the implication. The proof is by induction on the structure of $\phi$, where the $\Box$-depth $D(\phi)$ takes lexicographic precedence over formula structure (i.e., $\psi_1$ is less complex than $\psi_2$, if $D(\psi_1) < D(\psi_2)$, or if $D(\psi_1) = D(\psi_2)$ and $\psi_1$ is a subformula of $\psi_2$). The non-standard inductive cases are $[\phi]\psi$ and $\Box\psi$. In either case we only show one direction of the equivalence in the conclusion; the other direction is similar.
\paragraph*{Case $[\phi]\psi$}
Let $M_{s} \bisim N_{s'}$ and $M_s \models [\phi]\psi$. The latter is by definition equal to: $M_s \models \phi$ implies $M^\phi_s \models \psi$. Let us now assume $M_s \models \phi$.

First, from $M_{s} \bisim N_{s'}$ and $M_s \models \phi$, it follows by induction that $N_{s'} \models \phi$. 

Second, this not only holds for $s$ but for any $t$ in the domain of $M$ and $t'$ in the domain of $N$: from $M_{t} \bisim N_{t'}$ and $M_t \models \phi$, it follows by induction that $N_{t'} \models \phi$. 

This allows us to show that $M^\phi_s \bisim N^\phi_{s'}$, namely, given $\bisrel: M_s \bisim N_{s'}$, by the relation $\bisrel': M^\phi_{s} \bisim N^\phi_{s'}$ such that for all $t,t'$: $(t,t') \in \bisrel'$ iff ($(t,t') \in \bisrel$ and $M_t \models \phi$). We now show that the relation $\bisrel'$ is a bisimulation. The clause {\bf atoms} is obvious. For {\bf forth}, assuming some $(s,s')\in\bisrel'$, let $s \sim_a t$ in $M^\phi$. Let $t'$ be such that $(t,t') \in \bisrel$ and $s' \sim'_a t'$. From $(t,t') \in \bisrel$ and $M_t \models \phi$, we get with induction that $N_{t'} \models \phi$. Therefore, $(t,t') \in \bisrel'$. As $s' \sim'_a t'$ persists in $N^\phi$, the state $t'$ satisfies the requirements for {\bf forth}. The clause {\bf back} is shown similarly.

Third, having shown that $M^\phi_s \bisim N^\phi_{s'}$, and also using that $M_s \models [\phi]\psi$ and $M_s \models \phi$ implies $M^\phi_s \models \psi$, we now use the induction hypothesis for $\psi$ on pair of models $M^\phi_s$, $N^\phi_{s'}$, and thus obtain that $N^\phi_{s'} \models \psi$ as required. 

Winding up, we now have shown that ($M_s \models \phi$ implies $M^\phi_s \models \psi$) is equivalent to ($N_{s'} \models \phi$ implies $N^\phi_{s'} \models \psi$), i.e., $N_{s'} \models [\phi]\psi$, as required.


\paragraph*{Case $\Box\psi$}
Let $M_{s} \bisim N_{s'}$ and $M_s \models\Box\psi$. The latter is by definition equal to: for all $\phi_0 \in \langpl$,  $M_s \models[\phi_0]\psi$, i.e., for all $\phi_0 \in \langpl$, $M_s \models \phi_0$ implies $M_s^{\phi_0} \models \psi$. As $D(\phi_0) < D(\Box\psi)$ and $D(\psi) < D(\Box\psi)$, by twice using induction we obtain: for all $\phi_0 \in \langpl$, $N_{s'} \models \phi_0$ implies $N_{s'}^{\phi_0} \models \psi$, i.e., $N_{s'} \models \Box\psi$. 
\end{proof}
The property shown in the case announcement of the above proof is often used in the continuation and therefore highlighted in a corollary.
\begin{cor}
Let $\phi\in\langbapal$ such that $M_s \models \phi$. Then $M_{s} \bisim N_{s'}$ implies $M_{s}^\phi \bisim N_{s'}^\phi$.
\end{cor}
The next lemma may look obvious but is actually rather special: it holds for  \logicbapal{} but not, for example, for \logicapal{}, where the quantifiers are over formulas of arbitrarily large modal depth. Lemma~\ref{lemma.nequiv} plays a role in Section~\ref{sec.three} on expressivity.
\begin{lem} \label{lemma.nequiv}
Let two models $M_s,N_{s'}$ be given. If $M_s\bisim^n N_{s'}$, then $M_s\equiv^n N_{s'}$.
\end{lem}
\begin{proof}
We show the above by proving the following statement: \begin{quote} For all $n \in \Naturals$, for all $\phi \in \langbapal$ with $d(\phi) \leq n$, for all models $M_s,N_{s'}$: if $M_s\bisim^n N_{s'}$, then $M_s \models \phi$ iff $N_{s'}\models \phi$. \end{quote}
We prove this by refining the complexity measure used in the previous proposition: we now, additionally, give modal depth $d(\phi)$ lexicographic precedence over quantifier depth $D(\phi)$ (i.e., $\psi_1$ is less complex than $\psi_2$, if $d(\psi_1) < d(\psi_2)$, or if $d(\psi_1) = d(\psi_2)$ and $D(\psi_1) < D(\psi_2)$, or if $d(\psi_1) = d(\psi_2)$ and $D(\psi_1) = D(\psi_2)$ and $\psi_1$ is a subformula of $\psi_2$). For clarity we give the ---essentially different--- case $K_a \phi$ and also the ---essentially the same--- cases $[\phi]\psi$ and $\Box\psi$. The latter two apply to any $n \in \Naturals$ and do not require the induction over $n$. We let these cases therefore precede the case $K_a \phi$.


\paragraph*{Case $[\phi] \psi$}
Given are $M_s \bisim^n N_{s'}$ and $M_s \models [\phi]\psi$. We note that $d([\phi]\psi) = \max \{d(\phi),d(\psi)\}$ so that also $d(\phi), d(\psi) \leq n$. In this case of the proof we need to use induction on subformulas of $[\phi]\psi$. By definition, $M_s \models [\phi]\psi$ iff ($M_s \models \phi$ implies $M^\phi_s \models \psi$). 

In order to prove that $N_{s'} \models [\phi]\psi$, assume $N_{s'} \models \phi$. By induction, from $M_s \bisim^n N_{s'}$ and $N_{s'} \models \phi$ follows $M_s \models \phi$. From that and the given $M_s \models [\phi]\psi$ follows that $M^\phi_s \models \psi$. Similar to the proof of this inductive case of Lemma~\ref{bisimulation-preserves}, from $M_s \bisim^n N_{s'}$ follows $M^\phi_s \bisim^n N^\phi_{s'}$. From that and $M^\phi_s \models \psi$ follows $N^\phi_{t'} \models \psi$, as required.


\paragraph*{Case $\Box \psi$}
Let $M_{s} \bisim^n N_{s'}$ and $M_s \models\Box\psi$. As $d(\psi) = d(\Box\psi)$, we will now use that $D(\psi) < D(\Box\psi)$. This is therefore similar again to the same case in the previous Lemma~\ref{bisimulation-preserves}. By definition, $M_s \models\Box\psi$ is equal to: for all $\phi_0 \in \langpl$,  $M_s \models[\phi_0]\psi$, i.e., for all $\phi_0 \in \langpl$, $M_s \models \phi_0$ implies $M_s^{\phi_0} \models \psi$. It is now crucial to note that, as $\phi_0$ is Boolean, not only $D(\phi)=0$ but also $d(\phi_0)=0$. We therefore obtain by induction, as in the previous case $[\phi]\psi$ of this proof: for all $\phi_0 \in \langpl$, $N_{s'} \models \phi_0$ implies $N_{s'}^{\phi_0} \models \psi$, i.e., $N_{s'} \models \Box\psi$.


\paragraph*{Case $K_a \psi$}
Given are $M_s \bisim^{n+1} N_{s'}$ and $M_s \models K_a \phi$. Let now $t' \sim_a s'$. From $M_s \bisim^{n+1} N_{s'}$, $t' \sim_a s'$, and $n-\mathbf{back}$ follows that there is a $t \sim_a s$ such that $M_t \bisim^n N_{t'}$. From $M_s \models K_a \phi$ and $s \sim_a t$ follows that $M_t \models \phi$. As $d(\phi) = d(K_a\phi)-1 \leq n$, we can apply the induction hypothesis for $n$ and conclude that $N_{t'} \models \phi$. As $t'$ was arbitrary, $N_{s'} \models K_a \phi$.
\end{proof}
The interest of the above proof is the precedence of modal depth over quantifier depth, and of quantifier depth over subformula complexity. Essential in the proof is that in the case $\Box\psi$, for any $[\phi_0]\psi$ witnessing that, not only $D(\phi_0)=0$ but also $d(\phi_0) = 0$. Without $d(\phi_0) = 0$ the inductive hypothesis would not have applied. In contrast, the \logicapal\ quantifier is over formulas of arbitrary finite modal depth, also exceeding the modal depth of the initial given formula, which rules out use of induction.

Both for \logicapal\ and \logicbapal\ restricted bisimilarity does not imply restricted modal equivalence: \[ M_s \bisim_Q M'_{s'} \ \text{does not imply} \ M_s \equiv_Q M'_{s'}. \] The failure of this property is indirectly used in Proposition~\ref{prop.bapalmoreel} in the expressivity Section~\ref{sec.three}, later, for $Q = \{p\}$ (and for models with those same names).

\subsection{Arbitrary announcement normal form and necessity form}

We continue with results for the arbitrary announcement normal form and for the necessity form. The former are important to show decidability of \logicbapal, and the latter to show that the axiomatization is complete.

\begin{lem}\label{lem:aanf}
Every formula of $\langbapal$ is semantically equivalent to a formula in arbitrary announcement normal form.
\end{lem}

\begin{proof}
We give the proof by defining a truth preserving transformation $\delta$ from $\langbapal$ to $\langaanf$. 
This is defined with the following recursion:
$$
\begin{array}{lcllcl}
\delta(\atom) &=& \atom & 
  \delta(\neg\psi) &=& \neg\delta(\psi)\\
\delta(\psi\et\psi') &=& \delta(\psi)\et\delta(\psi')& 
  \delta(\knows_\agent\psi) &=& \knows_\agent\delta(\psi)\\ \ \\
\delta([\phi]\atom) &=& \delta(\phi\imp\atom) & 
  \delta([\phi]\neg\psi) &=& \delta(\phi\imp\neg[\phi]\psi)\\
\delta([\phi](\psi\et\psi')) &=& \delta([\phi]\psi\et[\phi]\psi') & 
  \delta([\phi]\knows_\agent\psi) &=& \delta(\phi\imp\knows_\agent[\phi]\psi)\\
\delta([\phi][\phi']\psi) &=& \delta([\phi\et[\phi]\phi']\psi) \hspace{1cm} & \\ \ \\
\delta(\Box\psi) &=& [\top]\Box\delta(\psi) & \delta([\phi]\Box\psi) &=& [\delta(\phi)]\Box\delta(\psi)
\end{array}
$$ 
We have to show that the translation is truth preserving and that the translation procedure terminates. 

The truth preservation is obvious for the clauses in rows 1 and 2 and for $\delta([\phi]\Box\psi) = [\delta(\phi)]\Box\delta(\psi)$. Concerning $\delta(\Box\psi) = [\top]\Box\delta(\psi)$ we note that $\Box\psi \eq [\top]\Box\psi$ is a valid equivalence. The translation clauses in rows 3 to 5 employ the valid equivalences $[\phi]\atom \eq (\phi\imp\atom)$, $[\phi]\neg\psi \eq (\phi\imp\neg[\phi]\psi)$, $[\phi](\psi\et\psi') \eq ([\phi]\psi\et[\phi]\psi')$, $[\phi]\knows_\agent\psi \eq (\phi\imp\knows_\agent[\phi]\psi)$, and $[\phi][\phi']\psi \eq [\phi\et[\phi]\phi']\psi$. These are well-known from \logicpal~\cite{plaza:1989,hvdetal.del:2007} (and for further reference we note that they are listed as axioms of \logicbapal\ in Section~\ref{sec.four}). 

We proceed by showing termination. For the clauses in rows 1, 2, and 6 we observe that each occurrence of $\delta$ on the right-hand side of an equation binds a formula that is less complex than the formula bound by $\delta$ on the left-hand side of the equation (and in $\delta(p)=p$, $\delta$ has disappeared on the right-hand side). For the clauses in rows 3 to 5 we refer to complexity measures in \logicpal~\cite{hvdetal.del:2007,balbianietal:2015}. Taking (e.g.) the measure~\cite[Def.~7.21]{hvdetal.del:2007}, \cite[Lemma~7.22]{hvdetal.del:2007} shows that the complexity of the right-hand side of each equivalence above is either equal to or lower than the complexity of the left-hand side of the equivalence. 
Therefore, $\delta$ will always return a formula in \langaanf.
\end{proof}

\begin{lem}\label{asdf}
Given are necessity form $\psi(\sharp)$, possibility form $\ov{\psi}\{\sharp\}$, and $\phi \in \langbapal$. Then $\neg\psi(\phi)$ is equivalent to $\ov{\psi}\{\neg\phi\}$.
\end{lem}
\begin{proof} 
This is easily shown by induction on the structure of necessity forms and using that $\neg(\chi \imp \psi(\phi))$ iff $\chi \et \neg\psi(\phi)$,  $\neg K_a \psi(\phi)$ iff $\M_a \neg\psi(\phi)$,   and $\neg [\chi]\psi(\phi)$ iff $\dia{\chi}\neg\psi(\phi)$.
\end{proof}

In the following lemma we show that a necessity form of arbitrary shape $\psi(\sharp)$ can be transformed into a necessity form of unique shape $\psi_1 \imp [\psi_2]\sharp$. With respect to instantiations $\psi(\theta)$ of such necessity forms this is a validity preserving transformation in both directions. Note that it is not a truth preserving transformation. This result will be used in Section~\ref{sec.four} to show that two versions of the axiomatization with different derivation rules are both complete (the translation is not only validity preserving but also derivability preserving).
\begin{lem} \label{lemma.tau}
Let $\psi(\sharp)$ be a necessity form and $\theta \in \langbapal$. Then there are $\psi_1,\psi_2\in\langbapal$ such that $\models\psi(\theta)$ iff $\models \psi_1 \imp [\psi_2]\theta$.
\end{lem}
\begin{proof}
Consider the following recursively defined translation $\tau$ (also employing subrecursion and subsubrecursion).
\[\begin{array}{lcl}
\tau(\sharp) & = & \top \imp [\top]\sharp \\ \ \\
\tau(\phi\imp\sharp) &=& \phi \imp [\top]\sharp \\
\tau(\phi \imp \psi_1 \imp \psi_2(\sharp)) &=& \tau(\phi\et\psi_1 \imp \psi_2(\sharp)) \\
\tau(\phi \imp K_a \psi(\sharp)) &=& \tau(\M_a\phi \imp \psi(\sharp)) \\
\tau(\phi \imp [\psi]\sharp) &=& \phi \imp [\psi]\sharp \\
\tau(\phi \imp [\psi_1](\psi_2 \imp \psi_3(\sharp))) &=& \tau(\phi\et[\psi_1]\psi_2 \imp [\psi_1]\psi_3(\sharp)) \\
\tau(\phi \imp [\psi_1]K_a\psi_2(\sharp)) &=& \tau(\M_a(\phi\et\psi_1) \imp [\psi_1]\psi_2(\sharp)) \\
\tau(\phi \imp [\psi_1][\psi_2]\psi_3(\sharp)) &=& \tau(\phi \imp [\psi_1\et[\psi_1]\psi_2]\psi_3(\sharp)) \\ \ \\
\tau(K_a\psi(\sharp)) & = & \tau(\psi(\sharp)) \\ \ \\
\tau([\phi]\sharp) &=& \top \imp [\phi]\sharp \\
\tau([\phi]K_a \psi(\sharp)) &=& \tau(\M_a\phi \imp [\phi]\psi(\sharp)) \\
\tau([\phi](\psi_1 \imp \psi_2(\sharp)) &=& \tau([\phi]\psi_1 \imp [\phi]\psi_2(\sharp)) \\
\tau([\phi][\psi_1]\psi_2(\sharp)) &=& \tau([\phi\et[\phi]\psi_1]\psi_2(\sharp))\end{array}\]
We can now observe that:
\begin{itemize}
\item {\em The translation $\tau$ terminates.} \\ Here we use again that formulas of shape $[\phi][\psi]\eta$ are at least as complex as $[\phi\et[\phi]\psi]\eta$. We should also observe that in any clause of shape $\tau(x \imp y) = \tau(z \imp w)$, $y$ is at least as complex as $w$, and that in any claus of shape $\tau([x]u) = \tau(z \imp w)$, $[x]u$ is at least as complex as $w$. The complexity the antecedent $z$ of the implication does not matter.
\item {\em $\tau(\psi(\theta))$ is a neccessity form of shape $\psi_1 \imp [\psi_2]\theta$, as required.} \\
There are four clauses in which $\tau$ does not appear on the right-hand side, and in all those cases the right-hand side has the required shape.
\item {\em For all $\theta\in\langbapal$, $\models\psi(\theta)$ iff $\models\tau(\psi(\theta))$, as required.} \\
This holds because all clauses are validity preserving in both directions. Apart from propositional tautologies this can be justified by one or more of the following observations:

In epistemic logic, $\models K_a \psi$ implies $\models \psi$, and $\models\psi$ implies $\models K_a \psi$. The validity of these in \logicbapal\ is also obvious. This is used in the, maybe surprising, case $\tau(K_a\psi(\sharp))$.

The \logicpal\ validities listed in the above Lemma~\ref{lem:aanf}, there justifying truth preservation, are also frequently used here; in addition (as this might otherwise be oblique) we use the validity $[\phi](\psi\imp\eta) \eq ([\phi]\psi\imp[\phi]\eta)$. This is used in various cases including $\tau([\phi](\psi_1 \imp \psi_2(\sharp))$.

In epistemic logic, $\models \phi \imp K_a \psi$ iff $\models \M_a \phi \imp \psi$. Let us prove this. We note that $\models \phi \imp K_a \psi$ implies $\models \M_a\phi \imp \M_aK_a \psi$, which, as $\M_a K_a\phi$ is equivalent in epistemic logic to $K_a \phi$, implies $\models \M_a\phi \imp K_a \psi$. From that and $\models K_a\psi \imp \psi$ we obtain $\models \M_a \phi \imp \psi$. For the other direction, $\models \M_a \phi \imp \psi$ implies $\models K_a\M_a \phi \imp K_a\psi$, which as $K_a \M_a \phi$ is equivalent in epistemic logic to $\M_a \phi$, implies $\models \M_a \phi \imp K_a\psi$. From that and $\models \phi \imp \M_a\phi$ we obtain $\models \phi \imp K_a \psi$. This is used in various cases including  $\tau(\phi \imp K_a \psi(\sharp))$.  \qedhere
\end{itemize} 
\end{proof}


\subsection{BAPAL validities involving the quantifier}

Some $\logicbapal$ validities are as follows. They play no auxiliary role as tools in later sections, but they serve to compare \logicbapal\ to other logics with quantification over information change, where such properties sometimes hold and sometimes not.

\begin{prop} \ 
\begin{itemize}
\item $\models \Box \phi \imp \phi$ (T)
\item $\models \Box \phi \imp \Box \Box \phi$ (4)
\item $\models \Box \Dia \phi \imp \Dia \Box \phi$ (MK)
\item $\models \Dia \Box \phi \imp \Box \Dia \phi$ (CR)
\end{itemize}  
\end{prop}
\begin{proof} \ 
\begin{itemize}
\item Let $M_s$ be arbitrary. Assume $M_s \models \Box\phi$. Then $M_s \models [\top]\phi$. As $\models [\top]\phi\eq\phi$, therefore $M_s \models \phi$.
\item The dual validity $\Dia\Dia \phi \imp \Dia \phi$ follows from the observation that for all $\phi_0,\psi_0\in\langpl$, $\dia{\phi_0}\dia{\psi_0}\phi$ is equivalent to $\dia{\phi_0\et\psi_0}\phi$.
\item The validity of MK is shown as in \logicapal. Assume that $M_s \models \Box\Dia\phi$. Given the set $\var(\phi)$ of propositional variables occurring in $\phi$, let ${\delta_s(\phi)}$ be the characteristic function for the valuation of $\var(\phi)$ in state $s$. We then have $M_s \models [{\delta_s(\phi)}] \Dia\phi$, and $M^{\delta_s(\phi)}_s \models \Dia \phi$. In the model $M^{\delta_s(\phi)}_s$ the valuation of the propositional variables in $\var(\phi)$ is constant in the domain (for each such $p \in \var(\phi)$, either $V(p) = \domain(M^{\delta_s(\phi)})$ or $V(p) = \emptyset$). We now use that for on models where the valuation of variables is constant, every formula that is true in a state of the model is also valid on the model. For \logicapal\ this result is found in~\cite[Lemma~3.2]{balbianietal:2008} (see also~\cite[Lemma~1]{hvdetal.theoria:2012}). The inductive proof of this result depends on the validity of $\Box\top \eq \top$ and $\Box\bot \eq \bot$. These are also valid for \logicbapal. From this it is easy to show that on such models, $\models \phi \imp \Box\phi$~\cite[Lemma~3.3]{balbianietal:2008}. Then, with duality and with the validity $\Box\phi\imp\phi$ shown in the first item, we obtain that $\Dia \phi \eq \phi$ and $\Box\phi \eq \phi$ are both valid on such models, and therefore also $\Dia\phi\imp\Box\phi$. From that and $M^{\delta_s(\phi)}_s \models \Dia \phi$ then follows that  $M^{\delta_s(\phi)}_s \models \Box\phi$. Thus $M_s\models \dia{\delta_s(\phi)}\Box\phi$, and $ M_s \models \Dia\Box\phi$.
\item In order to prove the validity of CR we have to show that: for all $\phi_0,\psi_0 \in \langpl$ and for all $M$ with non-empty denotation of $\phi_0$ and of $\psi_0$ there are $\phi'_0,\psi'_0\in\langpl$ such that $M^{\phi_0\phi'_0} \bisim M^{\psi_0\psi'_0}$. The obvious choice is $\phi'_0 = \psi_0$ and $\psi'_0 = \phi_0$, as \[ M^{\phi_0\phi'_0} = M^{\phi_0\psi_0} = M^{\phi_0\et\psi_0} = M^{\psi_0\et\phi_0} = M^{\psi_0\phi_0} = M^{\psi_0\psi'_0}. \qedhere \]
\end{itemize}
\end{proof}
The proof of the validity 4 is more direct than in~\cite{balbianietal:2008}, where it is used that $\dia{\chi}\dia{\psi}\phi$ is equivalent to $\dia{\dia{\chi}\psi}\phi$. Formula $\dia{\chi}\dia{\psi}\phi$ is not equivalent to $\dia{\chi\et\psi}\phi$ for all $\chi,\psi\in\langbapal$.

The proof of the validity CR for $\logicbapal$ is easier than for other quantified epistemic logics such as \logicapal, where the proof needs to `close the diamond at the bottom', which just as in the case of MK needs declaring the values of all atoms, i.e., we must choose $\phi'_0 = \psi'_0 = {\delta_s(\phi)}$. For such a proof see e.g.~\cite[Lemma~3.10]{hvdetal.papal:2020}.

Note that not all logics with quantification over announcements satisfy all the properties of the quantifier shown in this subsection. For example, the logic known as \logicpapal\ (quantifying over announcements of so-called positive formulas, corresponding to the universal fragment in first-order logic) does not satisfy the validity $\Box\phi\imp\Box\Box\phi$~\cite{hvdetal.papal:2020}.

\subsection{Boolean closure}

We will now define the novel notion of {\em Boolean closure}, and prove some lemmas for it. These are used when proving the soundness of the axiomatization of the logic $\logicbapal$, later.
\begin{defi}[Boolean closure]
Consider the union denoted $\ddot{P}$ of the set of atoms $P$ and the disjoint set of atoms $\{ p_{\phi_0} \mid \phi_0 \in \langpl(P)\}$. The {\em Boolean closure} of a model $M=(S, \sim, V)$ for atoms $P$ is the model $\ddot{M} = (S, \sim, \ddot{V})$ that is as $M$, except that the {\em Boolean closed valuation} $\ddot{V}$ is for atoms $\ddot{P}$ and such that $\ddot{V}(p) = V(p)$ for $p \in P$ and $\ddot{V}(p_{\phi_0}) = \I{\phi_0}_M$ for $p_{\phi_0} \in \ddot{P}\setminus P$. 
\end{defi}
As $P$ is countably infinite, and as the Booleans on $P$ can be enumerated, $\ddot{P}$ is also countably infinite. Given an epistemic model, then for each atom and for each Boolean there are also infinitely many atoms with the same value on the Boolean closure of that model. E.g., $p$ has the same value as $p_{p\et p}$ (i.e., the atom $q$ corresponding to the Boolean $p\et p$), and the same value as $p_{p\et p \et p}$ (the atom $q'$ corresponding to Boolean $p \et p \et p$), etc. We proceed with some other properties of the Boolean closure, in the form of lemmas. 
\begin{lem} \label{lemma.samevalue}
On a Boolean closed model $\ddot{M}$, for all Booleans, including Booleans of  atoms in $\ddot{P}\setminus P$, there is an atom with the same value.
\end{lem}
\begin{proof}
The proof is by induction on the structure of a Boolean $\phi\in\langpl(\ddot{P})$. 

If $\phi$ is an atom, it is obvious. 

If $\phi = \neg \psi$, by induction we may assume that there is an atom $p \in \ddot{P}$ such that $\ddot{V}(p) = \II{\psi}_{\ddot{M}}$. In case $p \in P$, then $p_{\neg p} \in \ddot{P}\setminus P$, so $\II{\neg \psi}_{\ddot{M}} = \II{\neg p}_{\ddot{M}} = \ddot{V}(p_{\neg p})$. In case $p \in \ddot{P}\setminus P$, there is a $\phi_0\in\langpl(P)$ such that $\II{\phi_0}_M = \ddot{V}(p)$. Now consider $q_{\neg\phi_0} \in \ddot{P}$. We then have that $\II{\neg\psi}_{\ddot{M}} = \II{\neg p}_{\ddot{M}} =  \II{\neg\phi_0}_M = \ddot{V}(q_{\neg\phi_0})$, as required.

If $\phi = \phi' \et \phi''$, by induction we may assume that there are $p',p'' \in \ddot{P}$ such that $\ddot{V}(p') = \II{\phi'}_{\ddot{M}}$ and $\ddot{V}(p'') = \II{\phi''}_{\ddot{M}}$, respectively. We need to distinguish various cases. If $p',p''\in P$, then $p_{p'\et p''} \in \ddot{P}$ and $\II{\phi'\et\phi''}_{\ddot{M}} = \ddot{V}(p_{p'\et p''})$. If $p',p''\in \ddot{P}\setminus{P}$, let $\psi',\psi''\in \langpl(P)$ be such that, respectively, $\ddot{V}(p') = \II{\psi'}_M$ and $\ddot{V}(p'') = \II{\psi''}_M$. Now consider $q_{\psi'\et\psi''} \in \ddot{P}$. As $\ddot{V}(q_{\psi'\et\psi''}) = V(\psi'\et\psi'')$, we now have that $\II{\phi'\et\phi''}_{\ddot{M}} = \II{p'\et p''}_{\ddot{M}} = \ddot{V}(q_{\psi'\et \psi''})$. The remaining two cases where $p'\in P$ and $p''\in \ddot{P}\setminus{P}$, and where $p'\in \ddot{P}\setminus{P}$ and $p''\in P$, are treated similarly.
\end{proof}
\begin{lem} \label{lemma.semclosure}
The semantics of $\Box$ on a Boolean closure are \[ \ddot{M}_s \models \Box \psi \ \ \text{iff} \ \ \mbox{for all } p \in \ddot{P} : \ddot{M}_s \models [p] \psi \]
\end{lem}
\begin{proof}
Let $\ddot{M}_s$ and $\psi \in \langbapal$ be given. By the semantic definition of $\Box\psi$, we have that $\ddot{M}_s\models\Box\psi$ iff $\ddot{M}_s \models [\phi_0]\psi$ for all $\phi_0 \in \langpl(\ddot{P})$. Assuming the latter, it follows that $\ddot{M}_s \models [p]\psi$ for all $p \in \ddot{P}$ because atoms are Booleans.

Let us now assume that $\ddot{M}_s \models [p]\psi$ for all $p \in \ddot{P}$. Towards a contradiction, let $\phi_0\in \langpl(\ddot{P})$ be such that $\ddot{M}_s \models \phi_0$ but $\ddot{M}^{\phi_0}_s \not \models \psi$. From  Lemma~\ref{lemma.samevalue} it follows that there is $q \in \ddot{P}$ such that $\ddot{V}(q) = \II{\phi_0}_{\ddot{M}}$. Therefore, $\ddot{M}_s \models q$ and $\ddot{M}^q_s \not \models \psi$. On the other hand, from $\ddot{M}_s \models q$ and the assumption that $\ddot{M}_s \models [p]\psi$ for all $p \in \ddot{P}$ we now obtain $\ddot{M}^{q}_s \models \psi$, a contradiction. Therefore $\ddot{M}_s \models \Box \psi$.
\end{proof}
The next lemma involves a translation $tr: \langbapal(\ddot{P}) \imp \langbapal(P)$ defined as $tr(p_{\phi_0}) = \phi_0$ for $p_{\phi_0} \in \ddot{P}\setminus P$ and all other clauses trivial. 
\begin{lem} \label{lemma.ii} Let $\psi \in \langbapal(\ddot{P})$, model $M = (S,\sim,V)$ for atoms $P$ and $s \in S$ be given. Then: \[ \ddot{M}_s \models \psi \ \text{iff} \ M_s \models tr(\psi). \] \end{lem} \begin{proof} The proof is by induction on $\psi$. As in other proofs in our contribution, it is important for the induction that the formula is declared before the model and the state in which it is interpreted, so that the induction hypothesis applied to a subformula, also applies to other models and states, namely in particular to restrictions of the given model and to any state in such a restriction. The interesting cases are:

\paragraph*{Case $\psi = q$} If $q \in P$, then $\ddot{M}_s \models q$ iff $M_s \models q$. If $q \in \ddot{P} \setminus P$, such that $q = p_{\phi_0}$ for some $\phi_0 \in \langpl(P)$, then $\ddot{M}_s \models q$ iff $M_s \models \phi_0$.  

\paragraph*{Case $\psi = [\psi']\psi''$}
\begin{flalign*}
&\ddot{M}_s \models [\psi']\psi'' \Eq \\
&\ddot{M}_s \models \psi' \text{ implies } \ddot{M}^{\psi'}_s \models \psi'' \Eq &&(*) \ \text{induction} \\
&M_s \models tr(\psi') \text{ implies } M^{tr(\psi')}_s \models tr(\psi'') \Eq \\
&M_s \models [tr(\psi')]tr(\psi'') \Eq \\
&M_s \models tr([\psi']\psi'')
\end{flalign*}

\paragraph*{Case $\psi = \Box \psi''$}
\begin{flalign*}
&\ddot{M}_s \models \Box \psi'' \Eq &&\text{by Lemma~\ref{lemma.semclosure}} \\
&\text{for all } q \in \ddot{P}, \ddot{M}_s \models [q] \psi'' \Eq \\
&\text{for all } \phi_0 \in \langpl(P), \ddot{M}_s \models [\phi_0] \psi'' \Eq \\
&\text{for all } \phi_0 \in \langpl(P), \ddot{M}_s \models \phi_0 \text{ implies } \ddot{M}_s^{\phi_0} \models \psi'' \Eq &&(**) \ \text{induction} \\ 
&\text{for all } \phi_0 \in \langpl(P), M_s \models \phi_0 \text{ implies } M_s^{\phi_0} \models tr(\psi'') \Eq \\
&\text{for all } \phi_0 \in \langpl(P), M_s \models [\phi_0] tr(\psi'') \Eq \\
&M_s \models tr(\Box \psi'').
\end{flalign*}

\medskip

\noindent $(*)$: By induction $\ddot{M}_s \models \psi'$ iff $M_s \models tr(\psi')$, and also $\ddot{(M^{tr(\psi')})}_s \models \psi''$ iff $M^{tr(\psi')}_s \models tr(\psi'')$, where it remains to show that $\ddot{(M^{tr(\psi')})} = \ddot{M}^{\psi'}$: the Boolean closure of the model restriction to $tr(\psi')$ is the model restriction to $\psi'$ of the Boolean closure. In order to show that, we first show show that $M^{tr(\psi')}$ and $\ddot{M}^{\psi'}$ have the same domain and accessibility relations, and then show that the valuation of atoms in the Boolean closure of $M^{tr(\psi')}$ corresponds to the valuation of atoms in $\ddot{M}^{\psi'}$. The models have the same domain because
\[ \II{\psi'}_{\ddot{M}} = \{ t \in S \mid \ddot{M}_t \models\psi' \}=_{\text{induction}} \{ t \in S \mid M_t \models tr(\psi') \} = \II{tr(\psi')}_M, \]
and they therefore also have the same accessibility relations $\sim_a$. The models obviously have the same valuation of atoms in $P$. Now consider $q_{\phi_0} \in \ddot{P}\setminus P$, where $\phi_0 \in \langpl(P)$. 

Let $V^{tr(\psi')}$ be the valuation of $M^{tr(\psi')}$. Then $\ddot{V^{tr(\psi')}}(q_{\phi_0}) = V^{tr(\psi')}(\phi_0) = \II{\phi_0}_M \inter \II{tr(\psi')}_M$. Let now $\ddot{V}^{\psi'}$ be the valuation of $\ddot{M}^{\psi'}$. Then $\ddot{V}^{\psi'}(q_{\phi_0}) = \ddot{V}(q_{\phi_0}) \inter \II{\psi'}_{\ddot{M}} = \II{\phi_0}_M \inter \II{\psi'}_{\ddot{M}}$. Again using that $\II{tr(\psi')}_M = \II{\psi'}_{\ddot{M}}$, we obtain that $\ddot{V^{tr(\psi')}}(q_{\phi_0}) = \ddot{V}^{\psi'}(q_{\phi_0})$, as required.

\medskip

\noindent $(**)$: We use the same argument as in $(*)$, only with (arbitrary) $\phi_0$ in the role of $\psi'$, where we note that $tr(\phi_0) = \phi_0$.
\end{proof}
The property shown in $(*)$ of the above proof is significant enough to be mentioned as a corollary.
\begin{cor} \label{cor.mr}
Let $\psi$ have non-empty denotation on a model $M$. The model restriction to $\psi$ of the Boolean closure of $M$ is the Boolean closure of the model restriction to $tr(\psi)$ of $M$: \[ \ddot{(M^{tr(\psi)})} = \ddot{M}^{\psi}. \]
\end{cor}
For $\psi \in \langbapal(P)$ we have that $tr(\psi)=\psi$, so that:
\begin{cor} \label{cor.one}
Let $\psi \in \langbapal(P)$ and model $M_s$ be given. Then $\ddot{M}_s \models \psi$ iff $M_s \models \psi$.\end{cor}

The following result will be needed to show the soundness of a rule in the axiomatization of \logicbapal. As a similar result from the literature (\cite[Prop.~3.7]{balbianietal:2008}) was later shown false, we give the proof in full detail. We recall that an atom $q$ is fresh with respect to $\phi$, if $\phi$ does not contain an occurrence of $q$.

\begin{lem} \label{anotherlemma}
Let $\psi\{\sharp\}$ be a possibility form, and $M = (S,\sim,V)$ an epistemic model. Then for all $\psi\{\Dia \phi\} \in \langbapal(P)$: $M_s\models \psi\{\Dia \phi\}$ iff $\ddot{M}_s\models\psi\{\dia{p} \phi\}$ for a fresh $p \in \ddot{P}\setminus P$.
\end{lem}
\begin{proof}
The proof is by induction on the structure of possibility forms. Note that in the formulation of the lemma the formula is declared prior to the model. Therefore, induction hypotheses for a subformula apply to model restrictions and states in those restrictions. 

For all cases, the direction from right to left is the direct application of the semantics of $\Box$, for the dual modality $\Dia$, and where we use that any $p \in \ddot{P}\setminus P$ must be such that $p = q_{\phi_0}$ for some $\phi_0 \in \langpl(P)$:

Let $\ddot{M}_s\models\psi\{\dia{p} \phi\}$ be given. From that, with Lemma~\ref{lemma.ii}, we get $M_s\models tr(\psi\{\dia{p} \phi\})$. As $p\in \ddot{P}\setminus{P}$, $tr(p) = \phi_0$ for some $\phi_0 \in \langpl(P)$, and therefore $M_s\models tr(\psi\{\dia{p} \phi\})$ equals $M_s\models \psi\{\dia{\phi_0} \phi\}$. By the (dual) semantics of $\Box$, from that we obtain $M_s \models \psi\{\Dia \phi\}$.

We continue with the direction from left to right.

\paragraph*{Case $\sharp$} Let $M_s \models \Dia \phi$. Then, there is a $\phi_0 \in \langpl$ such that $M_s \models \dia{\phi_0} \phi$.  Therefore $\ddot{M}_s \models \dia{p_{\phi_0}} \phi$. As $p_{\phi_0}\in\ddot{P}\setminus P$, $p_{\phi_0}$ is fresh with respect to $\phi\in\langbapal(P)$. 

\paragraph*{Case $\psi\et\psi'\{\sharp\}$} Let $M_s \models \psi \et \psi'\{\Dia\phi\}$. Then $M_s \models \psi$ and $M_s \models \psi'\{\Dia\phi\}$. From $M_s \models \psi$ and $tr(\psi)=\psi$ it follows from Lemma~\ref{lemma.ii} that $\ddot{M}_s \models \psi$. From $M_s \models \psi'\{\Dia\phi\}$ and induction it follows that $\ddot{M}_s \models \psi'\{\dia{p}\phi\}$, where $p \in \ddot{P}\setminus P$  is fresh. From $\ddot{M}_s \models \psi$ and $\ddot{M}_s \models \psi'\{\dia{p}\phi\}$ we now obtain $\ddot{M}_s \models \psi \et \psi'\{\dia{p}\phi\}$, where we observe that, as $p \in \ddot{P}\setminus P$ is fresh in $\psi'\{\dia{p}\phi\}$ and $\psi \in \langbapal(P)$, $p$ remains fresh in $\psi\et\psi'\{\dia{p}\phi\}$.

\paragraph*{Case $\M_a \psi'\{\sharp\}$} Let $M_s \models \M_a \psi'\{\Dia\phi\}$. Then there is $t \sim_a s$ such that $M_t \models \psi'\{\Dia\phi\}$. Therefore, by induction, there is $t \sim_a s$ such that $\ddot{M}_t \models \psi'\{\dia{p}\phi\}$ and where $p \in \ddot{P}\setminus P$  is fresh, so that $\ddot{M}_s \models \M_a \psi'\{\dia{p}\phi\}$.

\paragraph*{Case $\dia{\psi}\psi'\{\sharp\}$} Let $M_s \models \dia{\psi}\psi'\{\Dia\phi\}$. Then $M_s \models \psi$ and $M^\psi_s \models \psi'\{\Dia\phi\}$. Lemma~\ref{lemma.ii} and $tr(\psi)=\psi$ give that $M_s \models \psi$ iff $\ddot{M}_s \models \psi$. By induction we get that $M^\psi_s \models \psi'\{\Dia\phi\}$ iff $\ddot{M}^\psi_s \models \psi'\{\dia{p}\phi\}$, where $p \in \ddot{P}\setminus P$ is fresh, and where we have also used that $\ddot{M}^\psi = \ddot{(M^\psi)}$ (Corollary~\ref{cor.mr}). From $\ddot{M}_s \models \psi$ and $\ddot{M}^\psi_s \models \psi'\{\dia{p}\phi\}$ we now obtain that $\ddot{M}_s \models \dia{\psi}\psi'\{\dia{p}\phi\}$, where, just as in the case for conjunction, as $p$ is fresh in $\psi'\{\dia{p}\phi\}$, $p \in \ddot{P}\setminus P$ and $\psi \in \langbapal(P)$, $p$ remains fresh in $\dia{\psi}\psi'\{\dia{p}\phi\}$. 
\end{proof}

In the proof of the following Lemma~\ref{yetanotherlemma} (and only in this proof), we do not only have to keep track explicitly of the parameter set of atoms for which the logical language is defined, but similarly of the set of atoms for which a model is defined, and where this set of atoms may contain $P$ instead of being contained in it. We therefore resort to let $M_s(Q)$ denote that model $M$ is defined for variables $Q$ (and where $s$ is the designated state). This also implies that, given some $M_s(P)$, its Boolean closure is $\ddot{M}_s(\ddot{P})$. We also let $\logicbapal(Q)$ mean `\logicbapal\ for set of atoms $Q$'. 
Outside this proof, the parameter set of atoms remains $P$.

\begin{lem} \label{yetanotherlemma}
Let $\phi\in\langbapal(P)$ and $p \in \var(\phi)$, and let $M_s$ be an epistemic model. If $\phi$ is valid, then $\ddot{M}_s \models \phi[q/p]$ for any $q \in \ddot{P}\setminus P$.
\end{lem}

\begin{proof} 
  Let $\phi$ be $\logicbapal(P)$ valid. We first show that for any $Q$ with $P \subseteq Q$:
  \begin{equation}
    \label{eq:i}\text{If $\phi\in\langbapal(P)$ is $\logicbapal(P)$ valid, then $\phi$ is $\logicbapal(Q)$ valid.}
  \end{equation}
  It is more intuitive to show the contrapositive, where for notational convencience we replaced $\phi$ by $\neg\phi$:
  \begin{equation}
    \label{eq:ii}
    \text{If $\phi\in\langbapal(P)$ is $\logicbapal(Q)$ satisfiable, then $\phi$ is $\logicbapal(P)$ satisfiable.}
  \end{equation} 
This follows directly from the following statement:
\begin{equation}
  \label{eq:iii}
  \text{For all $\phi\in\langbapal(P)$ and for all $M_s(Q)$, $M_s(Q) \models \phi$ implies $M_s(P) \models \phi$.}
\end{equation} 
This statement is proved by formula induction. The non-trivial cases of the proof are $\phi = \dia{\phi''}\phi'$ and $\phi = \Dia\phi'$.

\paragraph*{Case $\phi = \dia{\phi''}\phi'$} Suppose $M_s(Q) \models \dia{\phi''}\phi'$. Then $M_s(Q) \models \phi''$ and $M^{\phi''}_s(Q) \models \phi'$. By induction it follows that 
$M_s(P) \models \phi''$ and $M^{\phi''}_s(P) \models \phi'$ (as usual, note that the inductive hypothesis applies to any model $N_t(Q)$, not merely to $M_s(Q)$; it therefore applies to $N_t = M^{\phi''}_s(Q)$). Therefore $M_s(P) \models \dia{\phi''}\phi'$.

\paragraph*{Case $\phi = \Dia\phi'$} Suppose $M_s(Q) \models \Dia\phi'$. Then there is $\psi_0 \in \langpl(Q)$ such that $M_s(Q) \models \dia{\psi_0}\phi'$, and therefore $M_s(Q) \models \psi_0$ and $M^{\psi_0}_s(Q) \models \phi'$. For any atom $q \in \var(\psi_0)$ such that $q \in Q\setminus P$, choose a fresh atom $p' \in P$ (fresh with respect to $\psi_0$ and $\phi'$, and with respect to prior choices of such atoms in $\var(\psi_0)$), and transform $M$ into $N = (S,\sim,V')$ with $V'(p')=V(q)$ and $V'(q) = V(p')$. Let $\psi'_0 \in \langpl(P)$ be the result of all such substitutions. It is clear that $M_s(Q) \models \psi_0$ iff $N_s(Q) \models \psi'_0$ and that $\II{\psi_0}_M = \II{\psi'_0}_N$. We also have $M^{\psi_0}_s(Q) \models \phi'$ iff $N^{\psi'_0}_s(Q) \models \phi'$, as the truth of $\phi'$ does not depend on swapping the values of variables $p'$ and $q$ not occurring in it. Note that $\phi'$ may contain quantifiers $\Dia$, so strictly this requires a subinduction on the number of $\Dia$ occurrences in $\phi'$, where in the inductive step, given a witness $\dia{\chi_0}$ for $\Dia$, we need to simultaneously substitute all occurrences of $q$ and $p'$ in $\chi_0$ for each other. 

So we now have $N_s(Q) \models \psi'_0$ and $N^{\psi'_0}_s(Q) \models \phi'$. By induction, and as $\psi'_0 \in \langbapal(P)$, $N_s(Q) \models \psi'_0$ implies $N_s(P) \models \psi'_0$, and $N^{\psi'_0}_s(Q) \models \phi'$ implies $N^{\psi'_0}_s(P) \models \phi'$.  From $N_s(P) \models \psi'_0$ and $N^{\psi'_0}_s(P) \models \phi'$ follows $N_s(P) \models \Dia\phi'$. Now observe that $N_s(P)$ and $M_s(P)$ only differ in variables not occurring in $\Dia\phi'$, so that $N_s(P) \models \Dia\phi'$ iff $M_s(P) \models \Dia\phi'$ (where similarly to above we take into account occurrences of $\Dia$ in $\phi'$). Therefore, $M_s(P) \models \Dia\phi'$.

This shows (\ref{eq:iii}). We can now quickly close the argument. Let $\phi$ be satisfiable for $Q$. Then there is $M_s(Q)$ such that $M_s(Q) \models \phi$. Using (\ref{eq:iii}), $M_s(P) \models \phi$. Therefore $\phi$ is satisfiable for $P$. This shows (\ref{eq:ii}). Therefore, we have now shown, for arbitrary $\phi$, (\ref{eq:i}): if $\phi$ is valid for $P$, and $P \subseteq Q$, then $\phi$ is valid for $Q$. We now apply (\ref{eq:i}) for $Q = \ddot{P}$.

Clearly $P \subset \ddot{P}$. So, if $\phi$ is $\logicbapal(P)$ valid, then $\phi$ is $\logicbapal(\ddot{P})$ valid. For any validity $\phi''$ in a logic $\logicbapal(P'')$, $p'' \in P''$, and fresh $q'' \in P''$, $\phi[q''/p'']$ is also a $\logicbapal(P'')$ validity. Therefore, given that $\phi$ is a validity of $\logicbapal(\ddot{P})$ and a (obviously) fresh $q \in \ddot{P}\setminus P$, also $\phi[q/p]$ is a validity of $\logicbapal(\ddot{P})$. Here is it important to observe that this is a validity for the class of epistemic models for variables $\ddot{P}$, and that this is not a validity for the class of Boolean closures of epistemic models for variables $P$. But of course, the latter is contained in the former: a Boolean closed model for $P$ is, after all, a model for $\ddot{P}$. Therefore, also $\ddot{M}_s(\ddot{P}) \models \phi[q/p]$, as required.
\end{proof}

\section{Expressivity} \label{sec.three}

Given logical languages $\lang$ and $\lang'$, and a class of models in which $\lang$ and $\lang'$ are both interpreted (employing a satisfaction relation $\models$ resp.\ $\models'$), we say that \emph{$\lang$ is at least as expressive as $\lang'$}, if for every formula $\phi'\in\lang'$ there is a formula $\phi\in\lang$ such that for all models $M$ and states $s \in \domain(M)$, $M_s \models'\phi'$ iff $M_s \models\phi$. If $\lang$ is not at least as expressive as $\lang'$ and $\lang'$ is not at least as expressive as $\lang$, then $\lang$ is \emph{incomparable} to $\lang'$. If $\lang$ is at least as expressive as $\lang'$, and $\lang'$ is at least as expressive as $\lang$, then $\lang$ is \emph{as expressive as} $\lang'$. If $\lang$ is at least as expressive as $\lang'$ but $\lang'$ is not at least as expressive as $\lang$, then $\lang$ is (strictly) {\em more expressive} than $\lang'$. The combination of a language with a semantics given a class of models determines a logic.  In this work we only consider model class $\mathcal{S}5$. We abbreviate ``given logic $L$ determined by language $\lang$, model class $\mathcal{S}5$ and satisfaction relation $\models$, and logic $L'$ determined by language $\lang'$, model class $\mathcal{S}5$ and satisfaction relation $\models'$, $\lang$ is at least as expressive as $\lang'$,'' by ``$L$ is at least as expressive as $L'$,'' and similarly for other expressivity terminology. Note that in this work the language $\langbapal$ is the same for \logicbapal\ and \logicapal, whereas the semantics of the quantifier are different in \logicbapal\ and \logicapal. We also consider language fragments, namely \langpal\ and \langel.

We show that \logicbapal{} is more expressive than \logicel{} and that \logicbapal{} is not as least as expressive as two other logics with quantification over announcements: \logicapal, and {\em group announcement logic} (\logicgal) \cite{agotnesetal.jal:2010}. It is not known whether \logicapal{} (or \logicgal{}) is at least as expressive as \logicbapal\ or not. We conjecture that it is not. To prove that, one would somehow have to show that the \logicbapal{}-$\Box$ in a given formula $\Box \phi$ can be `simulated' by an \logicapal{}-$\Box$ that is properly entrenched in preconditions and postconditions relative to $\phi$, thus providing an embedding of \langbapal{} into \langapal. This seems quite hard.

In the models depicted below we use the following visual conventions: the names of states are replaced by the sets of atoms true in those states; the accessibility relations for the two agents $a,b$ are reflexively and symmetrically (and transitively) closed, in other words, they partition the domain into equivalence classes; and the actual state (the designated world) is underlined.

\begin{prop} \label{prop.bapalmoreel}
 \logicbapal{} is {\em more} expressive than \logicel.
\end{prop}

\begin{proof}
To prove that \logicbapal{} is more expressive than \logicel{} we first observe that $\langel \subseteq \langbapal$ (and that on that restriction they have the same semantics), so that \logicbapal{} is at least as expressive as \logicel, and we then observe that the (standard) proof that \logicel{} is not at least as expressive as \logicapal{}~\cite{balbianietal:2008} can also be used to show that \logicel{} is not at least as expressive as \logicbapal{}. 

We recall the proof in~\cite{balbianietal:2008}, wherein the formula $\Dia (K_a p \et \neg K_b K_a p)$ is shown not to be equivalent to an epistemic logical formula $\psi$ as follows. There must be a propositional variable $q$ not occurring in $\psi$. Two models that are bisimilar except for $q$ will either make $\psi$ true in both or false in both. On the other hand, $\Dia (K_a p \et \neg K_b K_a p)$ may be true in one and false in the other, as it quantifies over variable $q$ as well. This quantification is \emph{implicit}, as $q \not \in \var(\Dia (K_a p \et \neg K_b K_a p)$. We can therefore easily make $\dia{q} (K_a p \et \neg K_b K_a p)$ true in one and false in the other, as shown below for $M_s$ and $M'_{s'}$. 

As the announcement $q$ witnessing the diamond is a  Boolean, this also proves the case for \logicbapal.

\bigskip

        \begin{tikzpicture}[>=stealth',shorten >=1pt,auto,node distance=5em,thick]
          \node (s) {\underline{$\{p\}$}};
          \node (sr) [right of=s] {$\{\}$};
          \node (bs) [below of=s] {\color{white} $\{p,q\}$};
          \node (bsr) [below of=sr] {\color{white} $\{q\}$};  
          \node (m) [left of=s] {$M_s$:};
          \draw  (s) edge node {$a$} (sr);
        \end{tikzpicture}
       \quad \quad \quad
        \begin{tikzpicture}[>=stealth',shorten >=1pt,auto,node distance=5em,thick]
          \node (s) {\underline{$\{p,q\}$}};
          \node (sr) [right of=s] {$\{\}$};
          \node (bs) [below of=s] {$\{p,q\}$};
          \node (bsr) [below of=sr] {$\{q\}$};
          \draw  (s) edge node {$a$} (sr);
          \draw  (bs) edge node {$a$} (bsr);
          \draw  (s) edge node {$b$} (bs);
          \draw  (sr) edge node {$b$} (bsr);
          \node (m) [left of=s] {$M'_{s'}$:};
        \end{tikzpicture}
\end{proof}

\begin{prop} \label{prop.notasap}
 \logicbapal{} is not at least as expressive as \logicapal.
\end{prop}
\begin{proof}
Consider (again, but to other usage)  \langapal{} formula $\Dia (K_a p \et \neg K_b K_a p)$. Let us suppose that there exists an equivalent \langbapal{} formula $\psi$. Given the modal depth $d(\psi)$ of $\psi$, consider two models $N_t, O_{t'}$, with a difference between them further away from the root than $d(\psi)$, ensuring that $N_t \bisim^{d(\psi)} O_{t'}$. 

Formally, $N_t$ and $O_{t'}$ can be defined as follows. Model $N_t$ has domain $\mathbb{Z}$, equivalence classes for relation $\sim_a$ consisting of pairs $\{2i,2i+1\}$ for $i \in \mathbb{Z}$ and for relation $\sim_b$ consisting of pairs $\{2i-1,2i\}$ for $i \in \mathbb{Z}$, and with $V(p) = \Union_{i \in \mathbb{Z}} \{4i-1,4i\}$. The actual state $t$ is state $0$. Note that $M_s \bisim N_t$, where $M_s$ is the model that was used in the previous proposition. Model $O_{t'}$ is as model $N_t$ (and with $t'=0$), except that the domain is restricted to the range $i \leq 4j$, where $j$ is the least positive integer for which $d(\psi) < 4j$ (so, on the left model $O$ is infinite, on the right it ends in two $p$ worlds). As the argument in the proof is abstract ($\psi$ is hypothetical) and only needs $j$ to be in excess of $\frac{d(\psi)}{4}$ (we only need to refer to formulas of modal depth larger than $d(\psi)$), the schematic visualization of these models below, wherein we have abstracted from the names of states, suffices in the proof.

\begin{figure}[ht]
\begin{tikzpicture}[>=stealth',shorten >=1pt,auto,node distance=5em,thick]
          \node (s) {\underline{$\{p\}$}};
          \node (sr) [right of=s] {$\{\}$};
          \node (srr) [right of=sr] {\color{white} $\{\}$};
          \node (sl) [left of=s] {\color{white} $\{p\}$};
          \node (sll) [left of=sl] {\color{white} $\{\}$};
          \node (slll) [left of=sll] {\color{white} $\{\}$};
          \node (st) [right of=srr] {\color{white} $\{p\}$};
         \node (stt) [right of=st] {\color{white} $\{p\}$};
            \draw  (s) edge node {$a$} (sr);
          \node (m) [left of=slll, node distance=1em] {$M_s$:};
        \end{tikzpicture}
\begin{tikzpicture}[>=stealth',shorten >=1pt,auto,node distance=5em,thick]

          \node (s) {\underline{$\{p\}$}};
          \node (sr) [right of=s] {$\{\}$};
          \node (srr) [right of=sr] {$\{\}$};
          \node (sl) [left of=s] {$\{p\}$};
          \node (sll) [left of=sl] {$\{\}$};
          \node (slll) [left of=sll] {\color{white} $\{\}$};
          \node (st) [right of=srr] {\color{white} $\{p\}$};
         \node (stt) [right of=st] {\color{white} $\{p\}$};
         \draw[dashed] (slll) edge (sll);
         \draw  (sll) edge node {$a$} (sl);
           \draw  (sl) edge node {$b$} (s);
           \draw  (s) edge node {$a$} (sr);
           \draw  (sr) edge node {$b$} (srr);
           \draw[dashed]  (srr) edge (st);
            \node (m) [left of=slll, node distance=1em] {$N_t$:};
       \end{tikzpicture}
     \end{figure}

\noindent \begin{tikzpicture}[>=stealth',shorten >=1pt,auto,node distance=5em,thick]

          \node (s) {\underline{$\{p\}$}};
          \node (sr) [right of=s] {$\{\}$};
          \node (srr) [right of=sr] {$\{\}$};
          \node (sl) [left of=s] {$\{p\}$};
          \node (sll) [left of=sl] {$\{\}$};
          \node (slll) [left of=sll] {\color{white} $\{\}$};
          \node (st) [right of=srr] {$\{p\}$};
         \node (stt) [right of=st] {$\{p\}$};
         \draw[dashed] (slll) edge (sll);
         \draw  (sll) edge node {$a$} (sl);
           \draw  (sl) edge node {$b$} (s);
           \draw  (s) edge node {$a$} (sr);
           \draw  (sr) edge node {$b$} (srr);
           \draw[dashed]  (srr) edge node {$> d(\psi)$} (st);
           \draw  (st) edge node {$b$} (stt);
           \node (m) [left of=slll, node distance=1em] {$O_{t'}$:};
       \end{tikzpicture}

\noindent \begin{tikzpicture}[>=stealth',shorten >=1pt,auto,node distance=5em,thick]

          \node (s) {\underline{$\{p\}$}};
          \node (sr) [right of=s] {\color{white} $\{\}$};
          \node (srr) [right of=sr] {\color{white} $\{\}$};
          \node (sl) [left of=s] {$\{p\}$};
          \node (sll) [left of=sl] {$\{\}$};
          \node (slll) [left of=sll] {\color{white} $\{\}$};
          \node (st) [right of=srr] {\color{white} $\{p\}$};
         \node (stt) [right of=st] {\color{white} $\{p\}$};
         \draw  (sll) edge node {$a$} (sl);
           \draw  (sl) edge node {$b$} (s);
           \node (m) [left of=slll, node distance=1em] {$O^\phi_{t'}$:};
       \end{tikzpicture}

\bigskip

We use Lemma~\ref{lemma.nequiv} that $n$-bisimilarity implies $n$-logical equivalence: from $N_t \bisim^{d(\psi)} O_{t'}$ it follows that $N_t \equiv^{d(\psi)} O_{t'}$ and thus, as the formula $\psi$ itself has depth $d(\psi)$, that $N_t \models \psi$ iff $O_{t'} \models \psi$ ($\dagger$). On the other hand, $N_t \not\models \Dia (K_a p \et \neg K_b K_a p)$ (obviously, consider the bisimilar $M_s$) whereas $O_{t'} \models \Dia (K_a p \et \neg K_b K_a p)$. To prove the latter we observe that any finite subset of the model $O$ can be distinguished from its complement by a formula in the logic (by `distinguished' we mean that the formula is true in all the states of that subset and false in all other states of the domain of that model), where we use that any state can be distinguished from all others by its distance to the rightmost terminal state\footnote{The rightmost state in $O$ is distinguished by $K_a p$. The state to its left (in the picture) is distinguished by $\M_b K_a p \et \neg K_a p$, and the state to the left of that by $\M_a \M_b K_a p \et \neg (\M_b K_a p \et \neg K_a p) \et \neg K_a p$. And so on. To distinguish any finite subset in $O$ from its complement, take the disjunction of the distinguishing formulas of its members.}, that is distinguished by $K_a p$. In particular, there must therefore be a distinguishing formula $\phi$ of a three-state subset of $O$ such that $O^\phi$ is as depicted. As $O^\phi_{t'} \models K_a p \et \neg K_b K_a p$, we get that $O_{t'} \models \dia{\phi} (K_a p \et \neg K_b K_a p)$, and thus $O_{t'} \models \Dia (K_a p \et \neg K_b K_a p)$. This is a contradiction with ($\dagger$). Therefore, no such $\psi \in \langbapal$ exists. 
\end{proof}
As a corollary of Proposition~\ref{prop.notasap} we can very similarly show that  \logicbapal{} is not at least as expressive as \logicgal, as in \logicgal{} we also quantify over announcements of arbitrarily large modal depth. We then use the same models as above but with an additional agent $c$ who has the identity accessibility relation on the model. On models where the accessibility relation for $c$ is the identity, $K_c \phi \eq \phi$ is valid for any $\phi$. The language of \logicgal\ has a primitive $\dia{c}\phi$ which stands for `there is a formula $\psi\in\langel$ such that $\dia{K_c\psi}\phi$' (see~\cite{agotnesetal.jal:2010}). On this three-agent model, $\dia{K_c\psi}\phi$ is equivalent to $\dia{\psi}\phi$, and as `there is a formula $\psi\in\langel$ such that $\dia{\psi}\phi$' is equivalent to `$\Dia\phi$' in the \logicapal\ semantics, we obtain that on this model $\dia{c}\phi$ is equivalent to $\Dia\phi$. We now copy the above argument but with $\dia{c} (K_a p \et \neg K_b K_a p)$ instead of $\Dia (K_a p \et \neg K_b K_a p)$. Therefore:

\begin{cor} \label{cor.gal}
 \logicbapal{} is not at least as expressive as \logicgal.
\end{cor}

\section{Axiomatization} \label{sec.four}

We now provide a sound and complete finitary axiomatisation for \logicbapal{}. 

\begin{defi}[Axiomatization \axiombapal]
The axiomatization \axiombapal{} of \logicbapal.
$$    \begin{array}{llll}
        {\bf P} & \text{propositional tautologies} &
        {\bf K} & K_a (\phi \imp \psi) \imp (K_a \phi \imp K_a \psi)\\
        {\bf T} & K_a \phi \imp \phi&
        {\bf 4} & K_a \phi \imp K_a K_a \phi\\
        {\bf 5} & \neg K_a \phi \imp K_a \neg K_a \phi&
        {\bf AP} & [\phi] p \eq (\phi \imp p)\\
        {\bf AN} & [\phi] \neg \psi \eq (\phi \imp \neg [\phi] \psi)&
        {\bf AC} & [\phi] (\psi \land \psi') \eq ([\phi] \psi \land [\phi] \psi')\\
        {\bf AK} & [\phi] K_a \psi \eq (\phi \imp K_a [\phi] \psi)&
        {\bf AA} & [\phi] [\psi] \psi' \eq [\phi \land [\phi] \psi] \psi'\\
        {\bf A\Box} & \Box\phi \imp [\psi_0] \phi \text{ where } \psi_0 \!\in\! \langpl &
        {\bf MP} & \phi \text{ and } \phi \imp \psi \text{ imply } \psi\\
        {\bf NecK} \!\! & \phi \text{ implies } K_a \phi&
        {\bf NecA} \!\! &  \phi \text{ implies } [\psi] \phi\\
        &&{\bf R\Box} & \psi \imp [\phi'][p]\phi \text{ for $p$ fresh implies } \psi \imp [\phi']\Box\phi 
\end{array}
    $$
\label{table.axiom}
\end{defi}
The rules and axioms in \axiombapal{} are as in the axiomatization of \logicapal, except for the axiom $\mathbf{A\Box}$ and the derivation rule $\mathbf{R\Box}$. A formula $\phi$ is a {\em theorem} (notation $\proves \phi$) if it belongs to the least set of formulas containing all axioms and closed under the derivation rules. 

The main interest of \axiombapal{} is that it is {\em finitary}, unlike other known axiomatizations for logics with quantification over announcements~\cite{balbianietal:2008,agotnesetal.jal:2010,GalimullinA17} (except for~\cite{BaltagOS18}, see Section~\ref{sec.one}). Essential towards proving that result is Lemma~\ref{anotherlemma}, stating that every diamond $\Dia$ in a possibility form is witnessed by the announcement $\dia{p}$ of a fresh variable\footnote{A similar lemma and finitary axiomatization reported for \logicapal{}~\cite{balbianietal:2008} are in fact incorrect, see \url{http://personal.us.es/hvd/errors.html}, although for \logicapal{} the infinitary axiomatization stands~\cite{balbianietal:2008,philippe.corrected:2015,balbianietal:2015}.}.

To demonstrate soundness and completeness of the axiomatization \axiombapal{} we can (still) use the line of reasoning in~\cite{balbianietal:2008}. 

We start with soundness, in other words, we will show that all axioms are validities and all rules are validity preserving (if all premisses of the rule are valid, then the conclusion is valid). The validity of axiom $\mathbf{A\Box}$ directly follows from the semantics of $\Box$. To establish the validity preservation of rule $\mathbf{R\Box}$, consider three versions of this derivation rule (let $\psi(\sharp)$ be a necessity form).
\begin{itemize}
\item $\mathbf{R\Box^\omega}$: ($\psi([\phi_0]\phi)$ for all $\phi_0\in\langpl$) implies $\psi(\Box\phi)$.
\item $\mathbf{R\Box^1}$: ($\psi([p]\phi)$ for a fresh $p \in P$) implies $\psi(\Box\phi)$.
\item $\mathbf{R\Box}$: ($\psi' \imp [\phi'][p]\phi$ for a fresh $p \in P$) implies $\psi' \imp [\phi']\Box\phi$.
\end{itemize}
We can analogously consider three axiomatizations:
\begin{itemize}
\item $\axiombapal^\omega = \axiombapal - \mathbf{R\Box} + \mathbf{R\Box^\omega}$
\item $\axiombapal^1 = \axiombapal - \mathbf{R\Box} + \mathbf{R\Box^1}$
\item \axiombapal
\end{itemize}
We show that all three of $\mathbf{R\Box^\omega}$, $\mathbf{R\Box^1}$, and $\mathbf{R\Box}$ are validity preserving, and that all three axiomatizations are sound and complete. It is thus a matter of taste which one is preferred. Note that \axiombapal$^\omega$ is infinitary whereas \axiombapal$^1$ and \axiombapal{} are finitary. Finitary axiomatizations are considered preferable over infinitary axiomatizations. Both $\mathbf{R\Box^1}$ and $\mathbf{R\Box}$ have (an instantiation of) a necessity form as premiss. The difference is that $\psi' \imp [\phi'][p]\phi$ is a particular necessity form whereas $\psi([p]\phi)$ can be any necessity form. As $\mathbf{R\Box}$ is more restrictive in logical structure, it may be considered preferable. Again, this is a matter of taste.

\begin{lem} \label{lemma.omega}
Derivation rule $\mathbf{R\Box^\omega}$ is validity preserving.
\end{lem}
\begin{proof}
This directly follows from the semantics of $\Box$. Let model $M_s$ be given. Assuming that $\psi([\phi_0]\phi)$ is valid for all $\phi_0\in\langpl$, we obtain that $M_s \models \psi([\phi_0]\phi)$ for all $\phi_0\in\langpl$. By the semantics of $\Box$, it follows that $M_s \models \psi(\Box\phi)$. As $M$ and $s \in \domain(M)$ were arbitrary, $\psi(\Box\phi)$ is valid.\end{proof}

\noindent We now show that the derivation rule $\mathbf{R\Box^1}$ is sound. This is the main technical result of this section, wherein we use results for the Boolean completion of models, introduced in Section~\ref{sec.two}.

\begin{prop}
Derivation rule $\mathbf{R\Box^1}$ is validity preserving.
\end{prop}
\begin{proof}
We recall that for any $\psi(\phi')\in\langbapal$ (where $\psi(\sharp)$ is a necessity form), $\overline{\psi}\{\neg\phi'\}$ (where $\psi'\{\sharp\}$ is a possibility form) is equivalent to $\neg\psi(\phi')$ (Lemma~\ref{asdf}).

Suppose that 
$\psi([p]\phi)\in \langbapal(P)$ is valid where $p$ is fresh. Let $M$ be any epistemic model and $s \in \domain(M)$.
Then from Lemma~\ref{yetanotherlemma} follows that $\ddot{M}_s \models \psi([q]\phi)$ for {\em any} fresh atom $q\in \ddot{P}\setminus P$. Therefore it is not the case that: there is a $q \in \ddot{P}\setminus P$ such that $\ddot{M}_s\models\overline{\psi}\{\dia{q} \neg\phi\}$ where $q$ is fresh. By applying the contrapositive of Lemma~\ref{anotherlemma}, we obtain $M_s\not\models\overline{\psi}\{\Dia \neg\phi\}$, i.e.,  $M_s\models\psi(\Box\phi)$.
\end{proof}

\begin{cor}
Derivation rule $\mathbf{R\Box}$ is validity preserving.
\end{cor}

\begin{thm}
\axiombapal{} is sound.
\end{thm}

We proceed to show that the three axiomatizations are complete, by way of showing that they all define the same set of theorems.

\begin{prop}[\cite{balbianietal:2015}]
$\axiombapal^\omega$ is sound and complete.
\end{prop}
\begin{proof} The soundness of the infinitary axiomatization $\axiombapal^\omega$ follows from the validity preservation of the infinitary derivation rule $\mathbf{R\Box}^\omega$ (Lemma~\ref{lemma.omega}). 

The completeness of $\axiombapal^\omega$ can be shown almost exactly as in~\cite{balbianietal:2015}, where we note that the original proof in~\cite{balbianietal:2008} contains errors that are corrected in this subsequent~\cite{balbianietal:2015}. The completeness part involves a canonical model construction and a fairly involved complexity measure and truth lemma with induction and subinduction using that complexity.  We will only point out the exact differences with~\cite{balbianietal:2015}. All proof details are therefore omitted.

The names of the axioms and rules in~\cite{balbianietal:2015} are different from ours. For example, the axiom $K_a(\phi\imp\psi) \imp (K_a\phi\imp K_a \psi)$ that we call {\bf K}, they call $A1$, etc. This is not a relevant difference. Instead of the axiom $\mathbf{A}\Box$ and the rule $\mathbf{R}\Box^\omega$ involving {\em Booleans}, the system in~\cite{balbianietal:2015} has an axiom and rule involving {\em epistemic formulas} (but that are otherwise identical): an axiom ``$\Box\phi \imp [\psi]\phi$ where $\psi\in\langel$'' and a rule ``$\psi([\phi']\phi)$ for all $\phi'\in\langel$, implies $\psi(\Box\phi)$''. Careful examination of the entire proof in~\cite{balbianietal:2015} reveals about a dozen occurrences of the word `epistemic' that have to be replaced by the word `Boolean' to match our case, where we can observe that they continue to perform the same role in the proof: their role is in all occasions that these formulas do not contain $\Box$ operators and thus have lower complexity, which is required for inductive assumptions. 

A final difference is that the system in~\cite{balbianietal:2015} contains an additional derivation rule ``$\phi$ implies $\Box\phi$'', that does not occur in $\axiombapal^\omega$. But it is derivable in the axiomatization of \logicapal\ in~\cite{balbianietal:2015} (and similarly in $\axiombapal^\omega$, by again substituting `Boolean' for `epistemic' in the following): let $\phi$ be given, then for all $\psi$ we get, with {\bf NecA}, $[\psi]\phi$, so in particular we get that for all \emph{epistemic} $\psi$, from which with their version of $\mathbf{R}\Box^\omega$ follows $\Box\phi$, as required.
\end{proof}

\begin{lem}[\cite{balbianietal:2008}] \label{lemma.trans}
A \axiombapal$^1$ derivation can be transformed into a  \axiombapal\ derivation. 
\end{lem}
\begin{proof}
This proof is found in~\cite{balbianietal:2008}, so we do not claim originality. However, the reader may appreciate that we present it in the context of Lemma~\ref{lemma.tau} and therefore in more detail.

First note that the translation $\tau$ defined in Lemma~\ref{lemma.tau} (page \pageref{lemma.tau}), where $\sharp$ is instantiated by a \langbapal\ formula, is not only validity preserving (in both directions) but also derivability preserving (in both directions). This we can justify as follows. Concerning the translation step $\tau(K_a\psi(\sharp)) = \tau(\psi(\sharp))$, we observe that $K_a \psi$ implies $\psi$ by the {\bf T} axiom of knowledge, whereas $\psi$ implies $K_a \psi$ by the {\bf NecK} derivation rule of necessitation for knowledge. All other translation steps correspond to derivable equivalences in \logicpal. 

Now consider a derivation in \axiombapal$^1$ and an application in that derivation of  rule $\mathbf{R\Box^1}$: ($\psi([p]\phi)$ for a fresh $p \in P$) implies $\psi(\Box\phi)$. From premiss $\psi([p]\phi)$ we first derive $\tau(\psi([p]\phi))$, we then apply derivation rule $\mathbf{R\Box}$ obtaining $\tau(\psi(\Box\phi))$, and from that we subsequently derive $\psi(\Box\phi)$. Successively doing this for all applications of $\mathbf{R\Box^1}$ in the \axiombapal$^1$ derivation transforms it into a \axiombapal\ derivation.
\end{proof}

With these results we can now easily demonstrate that not only $\axiombapal^\omega$ but also the other two axiomatizations are complete and define the same set of theorems. Below, let the name of the axiomatization stand for the set of derivable theorems. Again, we follow the same argument as in~\cite{balbianietal:2008}.
\begin{itemize}
\item $\axiombapal^\omega \subseteq \axiombapal^1$:  A derivation in $\axiombapal^\omega$ is not a finite sequence of formulas but a converse well-founded sequence of formulas, because a $R\Box^\omega$ rule application has an infinite number of premisses. We can transform such a $\axiombapal^\omega$ derivation into a $\axiombapal^1$ derivation as follows. If it contains no $R\Box^\omega$ rule applications it is already a $\axiombapal^1$ derivation. Otherwise, consider a $R\Box^\omega$ rule application with conclusion $\psi(\Box\phi)$. One of its infinite premisses must be $\psi([p]\phi)$ for a fresh atom $p$. Discarding all other premisses from that $R\Box^\omega$ rule application makes it a $R\Box^1$ rule application. Successively doing this for all $R\Box^\omega$ rule applications in the derivation (where we note that this is a finite number, as the derivation is converse well-founded) therefore transforms this $\axiombapal^\omega$ derivation into a $\axiombapal^1$ derivation.

\item $\axiombapal^1 = \axiombapal$: In Lemma~\ref{lemma.trans}, using the transformation $\tau$ defined in Lemma~\ref{lemma.tau}, was shown that a derivation with $R\Box^1$ rule applications can be transformed into one with $R\Box$ rule applications. This shows $\axiombapal^1 \subseteq \axiombapal$. The other direction of the mutual inclusion, $\axiombapal \subseteq \axiombapal^1$, is trivial, as a $R\Box$ rule application is also a $R\Box^1$ rule application, and therefore a \axiombapal\ derivation also a $\axiombapal^1$ derivation.

\item $\axiombapal \subseteq \axiombapal^\omega$: Here we use completeness of \axiombapal$^\omega$. Let a \axiombapal{} theorem be given. Using soundness of \axiombapal, we obtain that it is valid. The completeness proof of $\axiombapal^\omega$ involves showing that every $\axiombapal^\omega$ consistent formula is satisfiable. In other words, all \logicbapal{} validities are $\axiombapal^\omega$ theorems. So, our \axiombapal{} theorem, that is a \logicbapal{} validity, is a  $\axiombapal^\omega$ theorem.
\end{itemize}

\begin{thm}
\axiombapal{} is complete.
\end{thm}

\section{Conclusions and further research}
We proposed the logic \logicbapal{}. It is an extension of public announcement logic. It contains a modality $\Box \phi$ intuitively corresponding to: ``after every public announcement of a Boolean formula, $\phi$ is true''. We have shown that \logicbapal{} is more expressive than \logicel\ and not at least as expressive as \logicapal, and that it has a finitary complete axiomatization. 

For further research we wish to report the decidability of the satisfiability problem of  \logicbapal\ and of yet another logic with quantification over announcements, called {\em positive arbitrary public announcement logic}, \logicpapal. The logic \logicpapal{} has a primitive modality ``after every public announcement of a positive formula, $\phi$ is true''. The positive formulas correspond to the universal fragment in first-order logic. These are the formulas where negations do not bind modalities. It has been reported in~\cite{hvdetal.papal:2020}.

\section*{Acknowledgment}
  \noindent The authors wish to acknowledge the detailed anonymous LMCS reviews. Over the course of some revisions, we have enormously benefitted from the reviewers.

\bibliographystyle{alphaurl}
\bibliography{biblio2021}

\end{document}